\documentclass[twocolumn,showpacs,showkeys,preprintnumbers,amsmath,amssymb,prb,nofootinbib]{revtex4}
\usepackage{graphicx}
\usepackage{psfrag}
\def\td{\textup{d}}
\renewcommand\deg{\,^{\circ}}
\def\citep#1{[\onlinecite{#1}]}    
\def\cites#1{\cite{#1}}            
\def\litfig#1{Fig.~{#1}}           
\def\citefg#1{\litfig{\ref{#1}}}   
\def\eqref#1{(\ref{#1})}           
\def\secref#1{section \ref{#1}}    
\def\appref#1{appendix \ref{#1}}   
    
\def\real{\rm Re \ }
\def\imag{\rm Im \ }

\def\Lshift{\textup{\L}}       
\def\Tcdy{\widetilde{T}_c}     
\def\tcdy{\tilde{t}_c}

\def\Tnst{T^0_c}               

\begin{document}
\title{
High-frequency dynamical response of Abrikosov vortex lattice in flux-flow region
}
\author{F.~Pei-Jen Lin}
\email{fareh.lin@gmail.com}
\affiliation{NCTS, National Tsing Hua University, Hsinchu 300, Taiwan}
\author{Peter Matlock}
\email{pwm@induulge.net}
\affiliation{Research Department, Universal Analytics Inc., Airdrie, AB, Canada}

\begin{abstract}
The dynamical response of the Abrikosov vortex lattice in the presence
of an oscillating driving field is calculated by constructing an
analytical solution of the time-dependent Ginzburg-Landau equation.
The solution is steady-state, and work done by the input signal is
dissipated through vortex cores, mainly by scattering with phonons.
The response is nonlinear in the input signal, and is verified for
consistency within the theory.  The existence of well-defined
parameters to control nonlinear effects is important for any practical
application in electronics, and a normalised
distance from the normal-superconducting phase-transition boundary is 
found to be such a parameter to which the response is sensitive.
Favourable comparison
with NbN experimental data in the optical region is made, where the
effect is in the linear regime.  Predictions are put forward regarding
the suppression of heating and also the lattice configuration at high
frequency.
\end{abstract}
\maketitle
\section{Introduction}
Superconducting films are candidate substances for the improvement of
electronics technology in a myriad of applications.  While the low
resistance is very attractive in this regard, it has proved difficult
to control the nonlinear behaviour of such materials in response to
electromagnetic field\cites{O07}. When a magnetic field is strong enough to
penetrate into a superconductor in the form of quantised magnetic flux
tubes, the vortex state obtains as a mixed state of superconducting
phase punctuated by the vortices themselves.  Vortices are surrounded
by a supercurrent and can be forced into motion by the current
resulting from an applied field.
 
As a topological defect, a vortex is not only stable under 
perturbations\cites{Tinkham,Fg06} but cannot decay.  The
collection of vortices in a type-II superconductor forms what is
called vortex matter, and it is this which determines the physical
properties of the system rather then the underlying material
properties, in particular driving phase
transitions\citep{Tinkham,RL09}. In the mixed state, a superconductor
is not perfect; it exhibits neither perfect diamagnetism nor zero
electrical resistance.  The transport current ${\bf J}$ generates a
Lorentz force ${\bf F}=\Phi_0 {\bf J} \times \hat {\bf h}$ on the
vortex 
and forces it into motion, dissipating energy.

In reaching thermal equilibrium, energy is transferred via interactions 
between phonons and quasiparticle excitations.
Small-scale imperfections such as defects 
scatter the quasiparticles, affecting their dynamics.
In \emph{dirty} superconductors, impurities are plentiful and vortices experience a large friction.
This implies a fast momentum-relaxation process.
In contrast is the \emph{clean} limit, where impurities are rare and no such relaxation process is available.
It is in this situation of slow relaxation that the Hall effect appears.

Generally, the $H$-$T$ phase diagram\cites{RL09} of the vortex matter
has two phases. In the \emph{pinned phase} vortices are trapped by an
attractive potential due to the presence of large-scale defects, thus resistivity
vanishes. This phase contains what are known as glass states.
There is then
the \emph{unpinned phase} in which vortices can move when forced and
so a finite resistivity appears.  This phase is also known as the
flux-flow region and can be of two types. One type is a liquid state
where vortices can move independently; the other type is a solid state
in which vortices form a periodic Abrikosov lattice\cites{Ab57}
resulting from their long-range interacton.
One model for the transition between the pinned and unpinned phases appears in \citep{GB97}.

In the unpinned phase, 
the system is driven from equilibrium and experiences a
relaxation process. There are several ways to describe such a system.
A microscopic description\cites{Kpn02} invoking interactions between a
vortex and quasi-particle excitations at the vortex core provides a good
understanding of friction and sports good agreement with experiments
in the sparse-vortex region $H\ll H_{c2}$. There is also a macroscopic
description, the London approach, where vortices are treated either as
interacting point-like particles or an elastic manifold subject to a
pinning potential, driving force and friction\cites{CC91,GR66}. In the
small-field region, vortices behave as an array of elastic strings. 

In the dense-vortex region $H\gg H_{c1}$, where the magnetic field is
nearly homogeneous due to overlap between vortices, Ginzburg-Landau
(GL) theory, which describes the system as a field, provides a more
reasonable model.  
In dynamical cases, time-dependent GL
(TDGL) theory is appropriate\cites{HT71, TDGLnumerical,Kpn02}; in GL-type models,  
additional simplification can come from the lowest Landau level (LLL)
approximation which has proven to be successful in the vicinity of the
superconducting-normal (S-N) phase transition line $H\sim H_{c2}$.
This has been pursued in the static case\cites{RL09}  (without driving force)
and in the dynamic case with a time-independent transport
current\cites{LMR04}. It may be noted that in the glass state, zero
resistance within the LLL approximation cannot be
attained\cites{ZR07}.

Based on TDGL theory, we will study the dynamical response of a dense
vortex lattice forced into motion by an alternating current induced by
an external electromagnetic field.  
Vortices are considered which are free from being pinned and thermally 
excited, which in addition to thermal noise would produce entanglement and bending.
We assume the vortices can transfer work done by an external field to a heat bath.
Experimentally, a
low-temperature superconductor far away from the clean limit
is the best
candidate for attaining these conditions. We do not consider thermal
fluctuation effects specific to high-temperature superconductors.
In a dissipative system driven by a single-harmonic electric field
$E\cos\Omega\tau$, long after its saturation time we can expect the
system to have settled into steady-state behaviour, where the vortices are
vibrating periodically with some phase.  

The TDGL model in the presence of external electromagnetic field is
analysed and solved in \secref{models}.  The dynamical S-N
phase transition surface $\Tcdy(H,E,\Omega)$ is located in
$\{T,H,E,\Omega\}$-space. This surface coincides with the mean-field
upper-critical field $H_{c2}(T)$ in the absence of the applied field,
and with the phase-transition surface in the presence of the constant
driving field considered by Hu and Thompson\cites{HT71}.  We will
provide an analytical formalism for perturbative expansion in the
distance to $\Tcdy$, valid in the flux-flow region.  The
response of vortex matter forced into motion by the transport current
is studied in \secref{response}.  The current-density distribution 
and the motion of vortices are treated in \secref{motionofvortices}.  
In analysing the vortex lattice configuration in \secref{CML}, a 
method is utilised whereby the heat-generation rate is maximised.  
Next are discussed power dissipation, generation of higher harmonics, 
and the Hall effect.  An experimental comparison is made in 
\secref{Discussion} with Far-Infrared (FIR) measurement on NbN.  
Finally, some conclusions are made in \secref{Conclusion}.

\section{Flux-flow solution}
\label{models}
Let us consider a dense vortex system prepared by exposing a type-II
superconducting material to a constant external magnetic field 
${\bf H}=(0,0,-H)$ with magnitude $H_{c2}>H\gg H_{c1}$.
We also select the $c$ axis of the superconductor to be in the  $z$ direction. 
Let the superconductor carry an alternating electric current along the $y$ direction,
generated by an electric field $E(\tau)=E\cos\Omega\tau$ as shown in \citefg{twovortexfigure}.
Such a system when disturbed from its equilibrium state will undergo a
relaxation process. For our system, the TDGL equation\cites{HT71,KS98} is 
a useful extension of the equilibrium GL theory. 

In the dense-vortex region of the $H$-$T$ phase diagram, vortices
overlap and a homogeneous magnetic field obtains.  Describing the
response of such a system by a field, the order parameter $\Psi$ in 
the GL approach, is more suitable 
than describing vortices as
particle-like flux tubes, as is done in the London approach\cites{CC91}.

\subsection{Time-dependent Ginzburg-Landau model}

A strongly type-II superconductor is characterised by its large
penetration depth $\lambda$ and small coherence length $\xi$, 
$\kappa\equiv\lambda/\xi\gg1$. 
The difference between induced magnetic field and
external magnetic field is ${\bf H}-{\bf B}=-4\pi{\bf M}$.  In the
vicinity of the phase-transition line $H_{c2}(T)$ 
vortices overlap significantly, and ${\bf H}\approx{\bf B}$ making ${\bf M}$ small.
In this case, the magnetic field may be treated as homogeneous within the sample.
We will have in mind an experimental arrangement using a planar sample 
very thin compared with its lateral dimensions.
Since the characteristic length for inhomogeneity of electric field\cites{HT71}
$\xi_E^2=4\pi\lambda^2 \sigma_n/\gamma$ is then typically large compared with sample thickness, 
this implies that the electric field may also be treated as homogeneous throughout\cites{LMR04,HT71},
eliminating the need to consider Maxwell's equations explicitly. 

In equilibrium, the Gibbs free energy of the system is given by\cites{Tinkham}
\begin{eqnarray}
\label{GLfree} 
F[\Psi]&=&\int\td {\bf r} \bigg\{
\frac{\hbar^2}{2m_{ab}}|{\bf D}\Psi|^2
+\frac{\hbar^2}{2 m_c}|\partial_z \Psi|^2 \nonumber\\
&&{\ \ \ \ }-\alpha (\Tnst-T)|\Psi|^2+\frac{\beta}2 |\Psi|^4 \bigg\}
\end{eqnarray}
where $\Tnst$ is the critical temperature at zero field.
Covariant derivatives employed here preserve local
gauge symmetry and are two-dimensional;
$D_{\tau}=\partial_{\tau}+i\frac{e^*}{\hbar} A_{\tau}$
and ${\bf D}=\nabla^{(2)}-i \frac{e^*}{\hbar c} {\bf A}$.

Governing the dynamics of the field $\Psi$ is the 
TDGL equation
\begin{equation}
\label{1TDGL}
\frac{\hbar^2 \gamma}{2 m_{ab}} D_{\tau}\Psi =-\frac{\delta F}{\delta \Psi^*}
.\end{equation}
This determines the characteristic relaxation time of the order
parameter.  Microscopic derivation of TDGL can be found in
\citep{GE68,Kpn02} in which the values of $\alpha$, $\beta$ and $\gamma$
are studied. In the macroscopic case, these are viewed simply as
parameters of the model.  At microscopic scale, disorder is accounted
for by $\gamma$, the inverse of the diffusion constant; the relation of $\gamma$
to normal-state conductivity is discussed in \appref{Gamma}.

In standard fashion, ${\bf E}=-\nabla A_{\tau}-\frac1c\partial_{\tau} {\bf A}$
while ${\bf B}=\nabla \times {\bf A}$.
Our set of equations is completed\cites{Tinkham} by including Amp\`ere's law,
writing for the total current density
\begin{equation}
\label{Amplaw}
{\bf J}_0=\big(\frac c {4\pi}\big)\nabla \times \nabla \times {\bf A}= \sigma_n {\bf E}+{\mathcal J}_0
.\end{equation}
As we shortly make a rescaling of quantities, we have written $0$ 
subscripts here for clarity.  The first term is the normal-state  
conductivity. The second term can be written using a Maxwell-type equation
relating the vector-potential with the supercurrent,
\begin{eqnarray}
\label{current}
{\mathcal J}_0&=&-i \frac{\hbar e^*}{2m_{ab}}\left[
        \Psi^* {\bf D} \Psi
        -\Psi  ({\bf D} \Psi)^* 
        \right]
.\end{eqnarray}

This is a gauge-invariant model; we fix the gauge by 
considering the explicit vector potential
${\bf A({\bf r})}=(B y,0,0)$ and $A_{\tau}({\bf r},\tau)=y E\cos\Omega\tau$,
corresponding to an alternating transport current.
Each vortex lattice cell has exactly one fluxon.  We do not assume the electric field 
and the motion of vortices are in any particular direction relative 
to the vortex lattice, by way of rendering visible any anisotropy.

For convenience, we define some rescaled quantities.  The rescaled
temperature and magnetic field are $t=T/\Tnst$ and $b=H/H^0_{c2}$.  
$H^0_{c2}$ denotes the mean field upper-critical field, 
extrapolated from the $\Tnst$ region down to zero temperature.

In the $a$-$b$ plane of the crystal we make use of \emph{magnetic length}
$\xi_{\ell}$.  We define $\xi_{\ell}^2=\xi^2/b$ where
$\xi^2=\hbar^2/2m_{ab}\alpha \Tnst$.  The scale on the $c$-axis is
$\xi_c/\sqrt b$ with $\xi_c^2=\hbar^2/2m_c\alpha \Tnst$.
The coordinate anisotropy in $z$ is absorbed into this choice 
of normalisation, as can be seen in \eqref{Lop}.
The order parameter $\Psi$ is scaled by $\sqrt{2b \alpha \Tnst/\beta}$.  The time
scale is normalised as $\tau_s=\gamma\xi_{\ell}^2/2$.  Therefore,
frequency is $\omega=\Omega\tau_s$.
Note that $\omega$ is then inversely proportional to $b$.  The
amplitude of the external electric field is normalised with 
$E_0=2\hbar/e^*\xi^3\gamma$ so that $e=E/E_0$.

After our rescaling 
the TDGL equation takes the simple form
\begin{equation}
\label{2TDGL}
L\Psi-\frac1{2b}(1-t)\Psi+|\Psi |^2\Psi=0
\end{equation}
where the operator $L$ is defined as
\begin{equation}
\label{Lop}
L=D_{\tau}-\frac12{\bf D}^2-\frac12\partial_z^2
.\end{equation} 
With our specified vector potential, covariant derivatives are
$D_{\tau}=\partial_{\tau}+ivy\cos\omega\tau$, $D_x=\partial_x-iy$ and
$D_y=\partial_y$. We define $v=eb^{-3/2}$ for convenience. 
The TDGL equation is invariant under translation in $z$, thus 
the dependence of the solution in the $z$ direction can be
decoupled. 
$L$ is not hermitian;
\begin{equation}
L^{\dagger}=-D_{\tau}-\frac12{\bf D}^2-\frac12\partial_z^2
\end{equation}
where the conjugation is with respect to the usual inner product, defined below.

We will make extensive use of the Eigenfunctions of $L$ and $L^\dagger$ in what follows.
The Eigenvalue equation
\begin{equation}
L \varphi_{n,k_x} = \varepsilon_n \varphi_{n,k_x}
\end{equation}
defines the set of Eigenfunctions of $L$ appropriate for our analysis; 
this can be seen in \appref{LTDGL_sol}.
The convention is that 
$\varepsilon_n =\varepsilon_{n'}$
when and only when $n=n'$.
Taking corresponding\footnote{One `follows the sign' in front of the $D_{\tau}$ in
$L$ and switches it in the resulting $\varphi$ to get the `corresponding' 
Eigenfunction $\tilde{\varphi}$ for $L^\dagger$.}
Eigenfunctions of $L^\dagger$ to be $\tilde{\varphi}_{n,k_x}$, the orthonormality
$\langle \tilde{\varphi}_{n,k_x} , \varphi_{n',k'_x} \rangle = \delta_{nn'}\delta(k_x-k_x')$
may be chosen, so long as $\langle \tilde{\varphi}_{n,k_x} , \varphi_{n,k_x}\rangle \ne 0$.
Shown in \appref{LTDGL_sol}, crystal structure determines linear combinations of these
basis elements with respect to $k_x$; the resulting $\varphi_n$ functions are then
useful for expansion purposes below. 
The inner product is $\langle \tilde{\varphi}_m,\varphi_n\rangle = \langle \tilde{\varphi}_m^* \varphi_n\rangle$
where the brackets $\langle\cdots\rangle$ denote an integral\footnote{
Shortly we will be dealing with a periodic system, and we will normalise such integrations by the
unit cell volume and the period in time.} over space and time. 
To define averages over only time or space alone, 
we write $\langle\cdots\rangle_\tau$ or $\langle\cdots\rangle_{\bf r}$ respectively.
          \begin{figure}[ht] 
          \includegraphics[width=4cm]{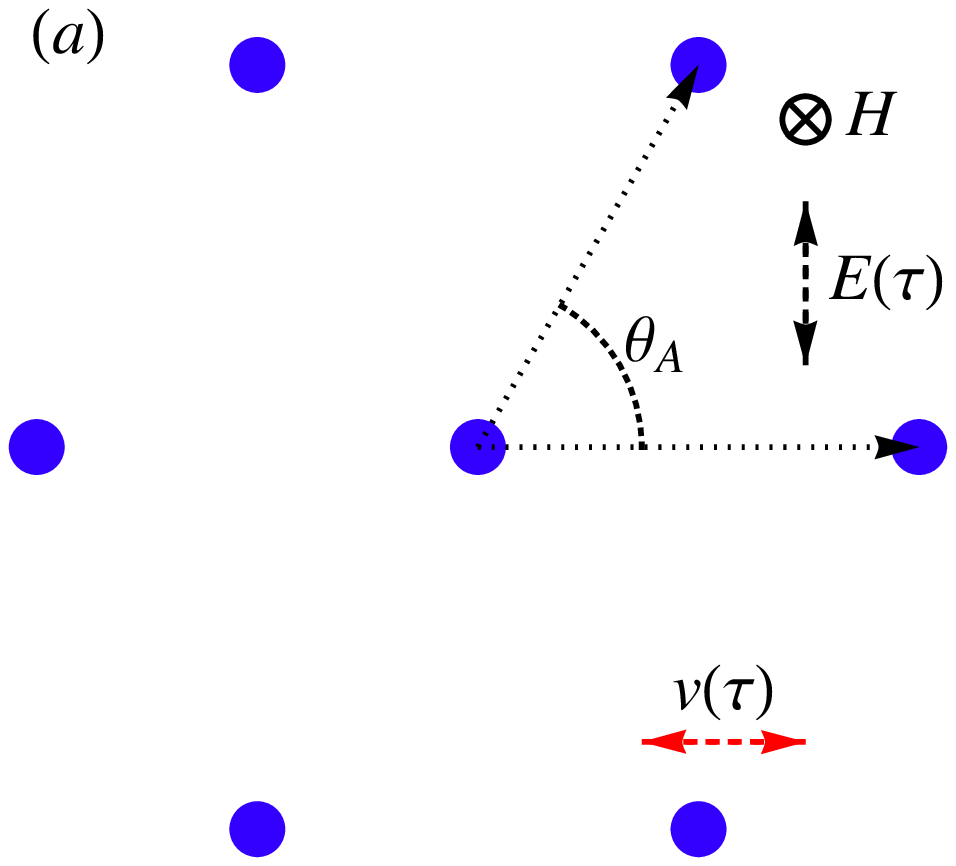}
	   \includegraphics[width=4cm]{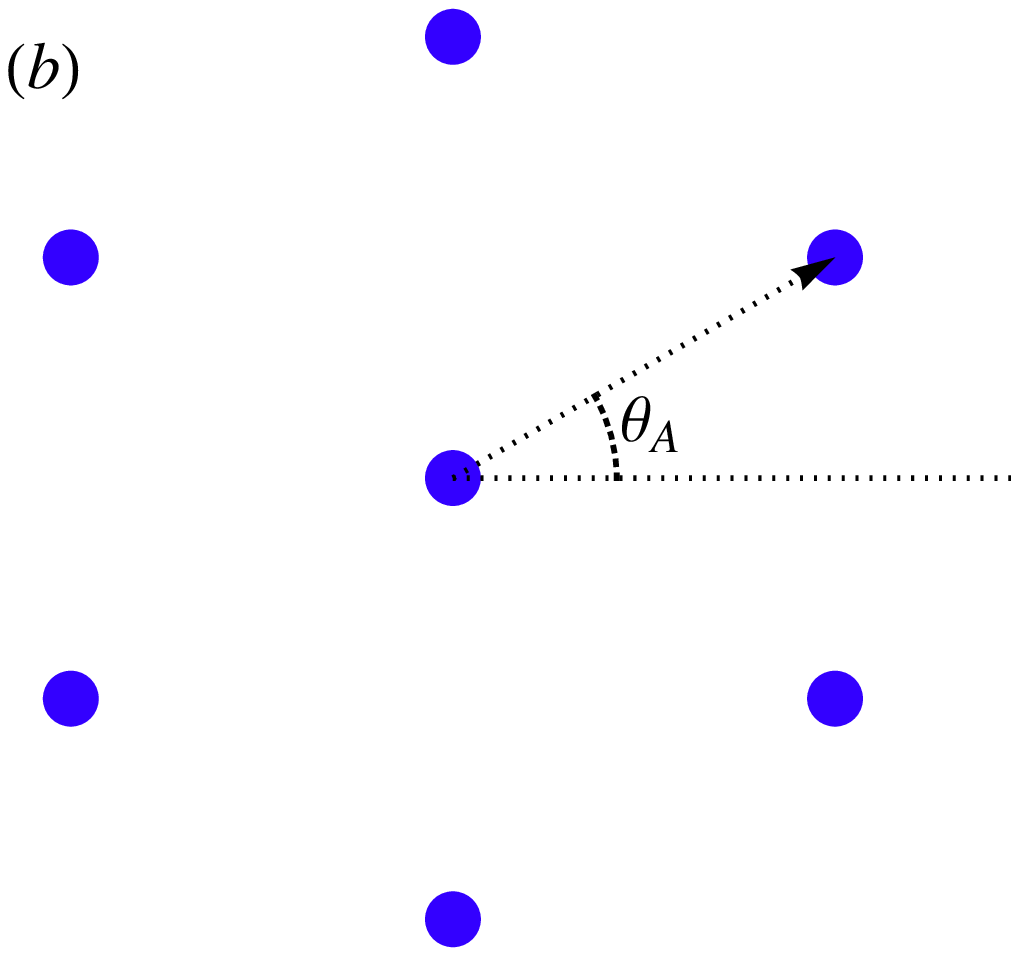}
        \caption{
	  Possible vortex lattice configurations:
        (a) Typical large angle, plotted for $\theta_A=60\deg$; 
        (b) Typical small angle, plotted for $\theta_A=30\deg$.
	 In the static case, both $\theta_A=60\deg$ and $\theta_A=30\deg$ 
	 are two particular angles will correspond to the same energy.
        The applied constant magnetic field ${\bf H}$ is along the $-z$ direction and
        time-dependent oscillating electric field ${\bf E}(\tau)$ is in the $y$ direction.
        $\theta_A$ is the the apex angle of the two defining lattice vectors.
        The two vectors for the rhombic vortex lattice are
        ${\bf a_1}=(\sqrt{2 \pi/\wp},0)$ and
        ${\bf a_2}=(\sqrt{2\pi/\wp}/2,\sqrt{2\pi\wp})$
        where $\wp=\frac12 \tan \theta_A$.
        The motion of vibrating vortices, indicated by the red arrow, is disccused in \secref{motionofvortices}.
        }
        \label{twovortexfigure}
        \end{figure}

\subsection{Solution of TDGL equation}
\label{TDGLsolution}
States of the system can be parametrised 
by $(t,b,e,\omega)$.  By changing temperature $t$, a system with some
fixed $(b,e,\omega)$ may experience a normal-superconducting phase 
transition as temperature passes below a critical value
$\tcdy(b,e,\omega)$.  Such a point of transition is also
known as a \emph{bifurcation point}.  

The material is said to be in the normal phase when $\Psi$ vanishes everywhere; 
otherwise the superconducting phase obtains, with $\Psi$ describing the vortex matter.
Because of the vortices, the resistivity in the superconducting phase need not be zero.
The S-N phase-transition boundary $\tcdy(b,e,\omega)$ separates the two phases.
To study the condensate, we will use a bifurcation expansion to solve \eqref{2TDGL}.  
We expand $\Psi$ in powers of distance from the phase transition boundary $\tcdy$.

\subsubsection{Dynamical phase-transition surface}
As in the static case, we can locate the dynamical phase-transition
boundary by means of the linearised TDGL equation\cites{Tinkham,KS98}.
This is because the order parameter vanishes at the phase transition,
and we do not need to consider the nonlinear term.  The linearised
TDGL equation is written
        \begin{equation}
        \label{LTDGL}
        L\Psi-\frac1{2b}(1-t) \Psi=0
        .\end{equation}

Of the Eigenvalues of $L$, only the smallest one $\varepsilon_{\bf 0}$,
corresponding to the highest superconducting temperature
$\tcdy$, has physical meaning. The S-N phase transition occurs
when the trajectory in parameter space intersects with the surface
        \begin{equation}
        \label{dyptline}
        \varepsilon_{\bf 0}-\frac1{2b}(1-t)=0
        \end{equation} 
where the lowest Eigenvalue is calculated in \appref{LTDGL_sol};
        \begin{eqnarray}
        \label{eigenvalue0}
        \varepsilon_{\bf 0}=\frac12+\frac{v^2}{4 (1+\omega^2)}
        .\end{eqnarray}
Utilising a $b$-independent frequency $\nu=b\omega$ and
amplitude $e$ of input signal, we write 
        \begin{equation}
        1-t-b=\frac{e^2/2}{b^2+\nu^2}
        .\end{equation}
In the absence of external driving field, $e=0$, the phase-transition
surface coincides with the well-known static-phase transition line
$1-t-b=0$ in the mean-field approach.  With time-independent electric
field at $\nu=0$, where the vortex lattice is driven by a fixed
direction of current flow, the dynamical phase-transition surface
coincides with that proposed in \citep{HT71}, but with a factor of
$1/2$.  This amplitude difference is familiar from elementary
comparisons of DC and AC circuits.

In the above equation, we can see that in the static case $e=0$,
the superconducting region is $1-t-b>0$.  In addition when $e\ne0$,
the superconducting region in the $b$-$t$ plane is smaller than the
corresponding region in the static case, as can be seen 
in \citefg{TwoPhaseTransitionProfile}(a). 

Finally, increasing frequency will increase the size of the
superconducting region, as in \citefg{TwoPhaseTransitionProfile}(b);
in the high-frequency limit, the area will reach its maximum, which is
the superconducing area from the static case. 
As with any damped system, response is diminished at higher frequencies.
	
The superconducting state does not survive at small magnetic field;
for example at $e=0.2$ in \citefg{TwoPhaseTransitionProfile}(a), the
material is in the normal state over most of the $H$-$T$ phase diagram. 
Later in this paper we will consider interpretation of this phenomenon.  
In particular, when discussing 
energy dissipation in \secref{powerlose}, we will see that the main contribution to the
dissipation is via the centre of the vortex core. At small magnetic
field, since there are fewer cores to dissipate the work done by the
electric field, the superconducting state is destroyed and the order
parameter vanishes.
        \begin{figure}[ht]
 	 \includegraphics[width=4.2cm]{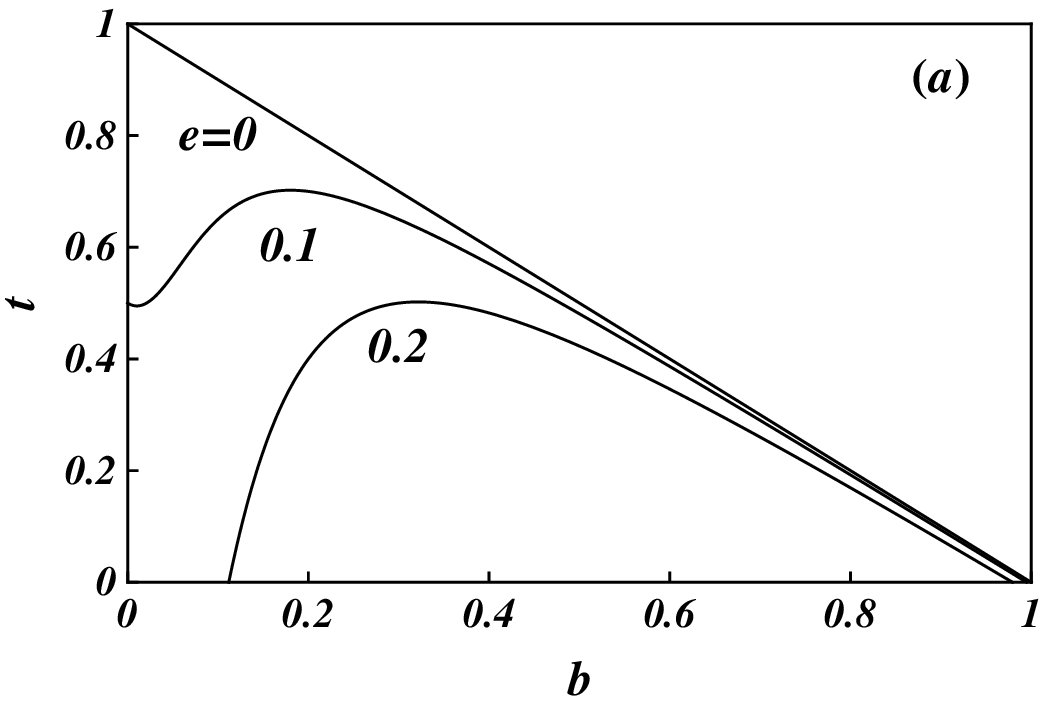}
        \includegraphics[width=4.2cm]{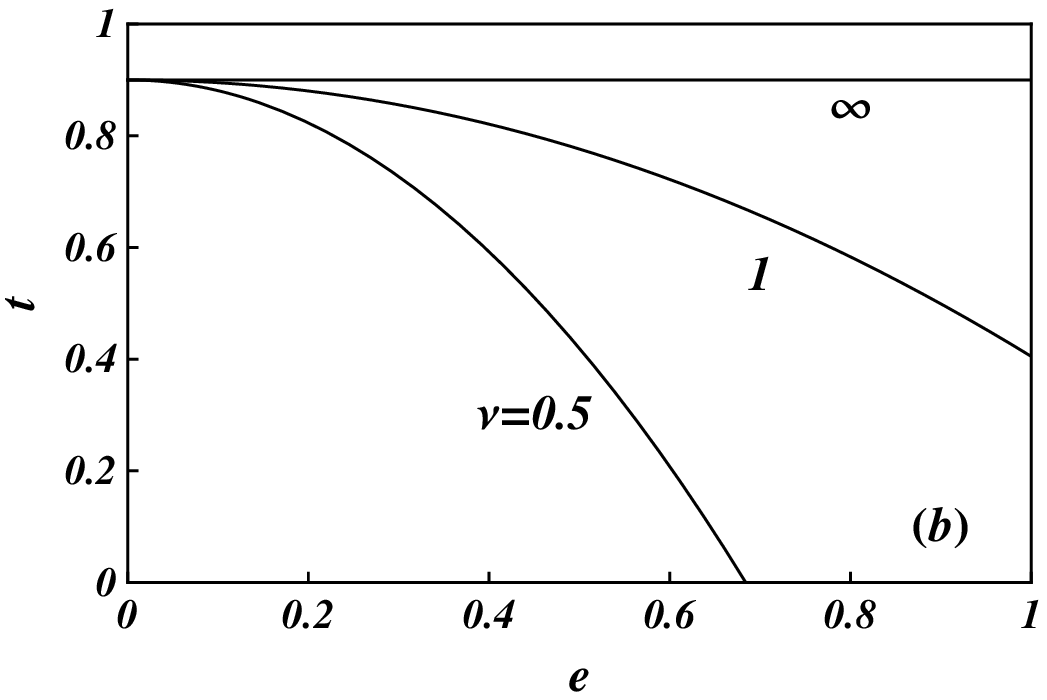}
        \caption{Dynamical superconducting-normal phase transition:
        (a) Critical temperature $\tcdy$ as a function of $b$ for various $e$ at $\nu=0.1$ and
        (b) $\tcdy$ as a function of $e$ for various $\nu$ at $b=0.1$.
        The straight line in (a) is the $e=0$ curve and corresponds to
        the mean-field phase-transition line $1-t-b=0$.
        States above each line are normal phase while
        the region below each line is superconducting. 
        $e$ suppresses the superconducting phase as shown in (a),
        while $\nu$ removes this suppression effect, as shown in (b).
        }
        \label{TwoPhaseTransitionProfile}
        \end{figure}
\subsubsection{Perturbative expansion}
\label{ptsol}
That the vortex matter dominates the physical properties of the system
is especially pronounced in the pinning-free flux-flow region.
Here we solve \eqref{2TDGL} by a bifurcation expansion\cites{L65,LMR04}.
Since the amplitude of the solution grows when the system departs from
the phase transition surface where $\Psi=0$, we can define a
distance from this surface as
    \begin{equation}
    \label{distance}
    \epsilon=\frac1{2b}(1-t)-\varepsilon_{\bf 0},
    \end{equation}
and expand $\Psi$ in $\epsilon$. The TDGL in terms of $\epsilon$ is
    \begin{equation}
    \label{shTDGL}
    \Lshift\Psi-\epsilon \Psi+ \Psi|\Psi|^2=0
    \end{equation}
where $\Lshift=L-\varepsilon_{\bf 0}$ is the operator $L$ shifted by its smallest Eigenvalue.
$\Psi$ 
is then written
    \begin{equation}
    \label{ansatz}
    \Psi=\sum_{i=0}^\infty  \epsilon^{i+1/2}   \Phi^{(i)}
    ,\end{equation}
and it is convenient to expand $\Phi^{(i)}$ in terms of our Eigenfunctions of $\Lshift$
    \begin{equation}
    \label{Phi_i}
    \Phi^{(i)}=\sum_{n=0}^\infty c^{(i)}_{n}\varphi_{n}   
    .\end{equation}

In principle, all coefficients $c^{(i)}_n$ in \eqref{ansatz} can be
obtained by using the orthogonal properties of the basis, which are explained in \appref{LTDGL_sol}.
Inserting $\Psi$ from equation \eqref{ansatz} into TDGL equation \eqref{shTDGL},
and collecting terms with the same order of $\epsilon$, we find that
for $i=0$
    \begin{equation}
    \label{eq1}
    \Lshift \Phi^{(0)}=0
    \end{equation}
and for $i=1$
    \begin{equation}
    \label{eq2}
    \Lshift\Phi^{(1)}-\Phi^{(0)}+\Phi^{(0)}|\Phi^{(0)}|^2=0
    .\end{equation}
For $i=2$
    \begin{equation}
    \label{eq3}
    \Lshift\Phi^{(2)}-\Phi^{(1)}+c^{(0)2}
        \big(
        2\Phi^{(1)}|\Phi_0|^2+\Phi^{(1)*}\Phi_0^2
        \big)=0
    \end{equation}
and so on.

Observing \eqref{eq1}, the solution for the equation is
    \begin{equation}
    \label{solution}
    \Phi^{(0)}=c^{(0)}_0 \varphi_0
    \end{equation}
where $\varphi_0$ is a particular linear combination of all Eigenfunctions with the smallest Eigenvalue.

The coefficient of $\epsilon^{1/2}$ can be obtained by calculating
the inner product of ${\tilde \varphi}_0$ with \eqref{eq2},
    \begin{equation}
    \label{c0}
    c^{(0)}_0=\frac1{\sqrt{\beta_{0}}}
    .\end{equation}
In the same way, the coefficient of the next order $\epsilon^{3/2}$,
can be obtained by finding the inner product of ${\tilde \varphi_0}$
with the $i=2$ equation \eqref{eq3},
    \begin{equation}
    \label{c10}
    c^{(1)}_0=\frac1{2 \beta_0 }\sum_{n=1}^{\infty} \big(
            2 c^{(1)}_n \langle {\tilde \varphi_0},\varphi_n |\varphi_0|^2\rangle
            +c^{(1)*}_n \langle{\tilde\varphi_0},\varphi_n^* \varphi_0^2\rangle
            \big)
    .\end{equation}
The inner product of ${\tilde \varphi}_m$ on \eqref{eq2}
gives the coefficient for $m>0$
    \begin{equation}
    \label{c1m}
    c^{(1)}_m=-\frac{\beta_m/\beta_0}{\varepsilon_m-\varepsilon_0}
    ,\end{equation}
and
    \begin{equation}
    \label{betam}
    \beta_{m}\equiv \langle {\tilde \varphi_m},\varphi_0|\varphi_0|^2\rangle
    .\end{equation}
The solution of TDGL is then
    \begin{equation}
    \label{solall}
    \Psi=\epsilon^{1/2} \frac{\varphi_0}{\sqrt{\beta_0}}
    +\epsilon^{3/2}\sum_{n=0}^\infty c^{(1)}_n \varphi_n
    +{\mathcal O} (\epsilon^{5/2})
    .\end{equation}
  
In this paper we will restrict
our discussion to the region near $\tcdy$ where the
next-order correction can be disregarded;
    \begin{equation}
    \label{sol}
    \Psi\approx\sqrt{\frac{\epsilon} {\beta_0}} \varphi_0
    .\end{equation}

We would like to emphasise that our discussion at this order
is valid in the vicinity of the phase-transition boundary and in particular 
for a superconducting system without vortex pinning.  In such a system,
vortices move in a viscous way, resulting in flux-flow resistivity; 
no divergence of conductivity is expected. 
Our results based on \eqref{sol} were calculated 
at $\epsilon^{1/2}$ order, where only the lowest 
eigenvalue $n=0$ of the TDGL operator $L$ makes an appearance.  

The next-order correction is at order $\epsilon^{3/2}$, and there is now a contribution
from higher Landau levels.  From the symmetry argument in 
\citep{LR99,L65}, as long as the hexagonal lattice remains the stable
configuration for the system, the next-order contribution comes from
the sixth Landau level with a factor $(\varepsilon_6-\varepsilon_0)^{-1}$.
Even in the putative case of a lattice deformed slightly away from
a hexagonal configuration, the next contributing term is $n=2$,
since in our system the lattice will remain rhombic.
 
\subsection{Vortex-lattice solution} 
\label{MLS}

The vortex lattice has been experimentally observed since the 1960s
and its long-range correlations have been clearly observed\cites{Kim99} 
with dislocation fraction of the order $10^{-5}$.
Remarkably, the same techniques can be used to study the 
structure and orientation of moving vortex lattice with steady current\cites{Fg02}, 
and with alternating current in the small-frequency
regime\cites{Fg06}.  In this subsection, we will discuss the
configuration of the vortex lattice in the presence of alternating
transport current in the long-time limit.

In the dynamical case, the presence of an electric field breaks the
rotational symmetry of an effectively isotropic system to the discrete
symmetry $y \rightarrow -y$.  In contrast, a rhombic lattice preserves
at least a symmetry of this kind along two axes, and the special case
of a hexagonal lattice preserves sixfold symmetry.

The area of a vortex cell is determined by the quantised flux in the 
vortex, which is $2\pi$ in terms of our rescaled variables.
As shown in \citefg{twovortexfigure}, we choose a unit cell $C$ defined 
by two elementary vectors ${\bf a}_1$ and ${\bf a}_2$. We will first
construct a solution for an arbitrary rhombic lattice parameterised by
an apex angle $\theta_A$.

Consideration of translational symmetry in the $x$ direction leads to
the discrete parameter $k_l=2\pi l/a_1=\sqrt{2 \pi \wp}l$.
In \appref{LTDGL_sol} we show that in the long-time limit the
lowest-eigenvalue steady-state Eigenfunctions of $L$ must therefore combine to form
    \begin{equation}
    \label{rhossol}
        \varphi_0=\sqrt[4]{2 \sigma}\sum_{l=-\infty}^{\infty}
        e^{i \frac{\pi}2 l(l-1)}
        e^{i k_l (x- v \sin\omega \tau/\omega)} u_{k_l}(y,\tau)
    .\end{equation}
Here $\varphi_0$ is normalised as
    \begin{eqnarray}
    \label{normalised}
    \langle\tilde{\varphi}_0,\varphi_0\rangle \nonumber\equiv1
    .\end{eqnarray}
The function $u$ is given by
    \begin{eqnarray}
    \label{ukv}
    u_{k_l}(y,\tau)=c(\tau)e^{  -\frac12
               \big[
                    y-k_l+i {\tilde v} \cos(\omega \tau-\theta)
                \big]^2
                }
    ,\end{eqnarray}
with
    \begin{eqnarray}
    \label{ct}
    c(\tau)=e^{
            -\frac{{\tilde v}^2}{4}
                \big[
                    \sin^2\theta+
                    \cos 2(\omega \tau-\theta)+
                    \frac1{2\omega}\sin2(\omega \tau-\theta)
                \big]
            }
	.\end{eqnarray}
In analogy with a forced vibrating system in mechanics, a
phase $\theta=\tan^{-1} \omega$ and a reduced velocity $\tilde v=v\cos\theta$ 
have been introduced for convenience in \eqref{ct}.  The zero electric-field 
limit, large-frequecy limit and zero frequency limit are consistent 
with previous studies concluded in \appref{LTDGL_sol}.

In our approximation, the $\beta_0$ in \eqref{sol} is a
time-independent quantity from \eqref{betam} and
    \begin{eqnarray}
    \label{beta0}
    \beta_0&=&
    \frac{\sqrt{\sigma}}{2 \pi}
    \int_0^{2\pi}\td (\omega\tau)
    \bigg\{
    e^{\frac{v_{\omega}^2}{4}
            \big( 1+
                \cos2\omega\tau
                +(\omega-1/\omega)\sin2\omega\tau
                \big)
        }
         \nonumber\\
    &&
     \sum_{p=-\infty}^{\infty}
        e^{- \frac12(
        k_p-i v_{\omega} \cos \omega\tau
        )^2
        } \nonumber\\
&&     \sum_{q=-\infty}^{\infty} (-)^{pq}
        e^{-\frac12(
        k_q-iv_{\omega} \cos \omega\tau
    )^2\big)
    }
    \bigg\}
    .\end{eqnarray}
where $v_{\omega}=v/(1+\omega^2)$.

In the small signal limit $v\rightarrow0$, $\beta_0$ reduces to the
Abrikosov constant.  The Abrikosov constant with either
$\theta_A=30\deg$ or $60\deg$ minimises the GL free energy
\eqref{GLfree} in the static state\cites{L65}.
To be more explicit, $\beta_0$ can be expanded in terms of the amplitude of input signal. 
In powers of $v$,
    \begin{eqnarray}
    \label{expbeta0}
    \frac{\beta_0}{\beta_A}=1+\frac12v_{\omega}^2\big(\frac32-
    \frac{\beta_b}{\beta_A}\big)+\mathcal O(v^3)
    \end{eqnarray}
and we find it convenient to write in terms of $v_{\omega}$.
The first term in $\beta_0$ is the Abrikosov constant
    \begin{equation}
    \beta_A=\sqrt{\sigma}\sum_{p,q=-\infty}^{\infty}(-)^{pq}e^{-\frac12(k_p^2+k_q^2)}
    .\end{equation}
For hexagonal lattices $\beta_A\sim1.159$, whereas for a square lattice $\beta_A\sim1.18$.
The next term in $\beta_0$ is $v_{\omega}^2$ with a coefficient
    \begin{equation}
    \beta_b=\frac{\sqrt{\sigma}}2\sum_{p,q=-\infty}^{\infty}(-)^{pq}
        (k_p^2+k_q^2)e^{-\frac12(k_p^2+k_q^2)}
    .\end{equation}
We see that at high frequency, the correction in higher order terms 
of $v_{\omega}$ can be disregarded.

\section{Response} 
\label{response}
In this section we discuss the current distribution and motion of vortices,
energy transformation of the work done on the system into heat,
nonlinear response and finally the Hall effect.
\subsection{Motion of vortices} 
\label{motionofvortices}

In addition to the conventional conductivity attributable to the
normal state, there is an overwhelming contribution due to the
superconducting condensate in the flux-flow regime, tempered only by
the dissipative properties of the vortex matter.  In this section we
will examine the supercurrent density to investigate the motion of the
vortex lattice. We consider a hexagonal lattice in a fully dissipative system;
the non-dissipative part known as the Hall effect will be discussed in 
\secref{halleffect}.

The supercurrent density ${\mathcal J}({\bf r},\tau)$ is obtained
by substitution of the solution \eqref{sol} into \eqref{current}.
    \begin{eqnarray}
    \label{jx}
    {\mathcal J}_x=
    \frac {\epsilon}{\beta_0}
    \sum_{p,q=-\infty}^{\infty}
    \bigg( \frac{k_p+k_q}2-y\bigg) g_{p,q}({\bf r},\tau)
    ,\end{eqnarray}
and
    \begin{eqnarray}
    \label{jy}
    {\mathcal J}_y=
    \frac {\epsilon}{\beta_0}
    \sum_{p,q=-\infty}^{\infty}
            \bigg(
                \frac{i (k_p-k_q)}{2}-{\tilde v}\cos(\omega \tau -\theta)
            \bigg)
            g_{p,q}
    \end{eqnarray}
where 
    \begin{eqnarray}
    g_{p,q}=e^{-i\frac{\pi}2(p^2-q^2-p+q)}
    e^{i(k_q-k_p)(x-\frac{v}{\omega} \sin\omega \tau)}
    u^*_{k_p}u_{k_q}\nonumber
    ,\end{eqnarray} 
and $u$ is given in \eqref{ukv}. 
Observing \citefg{CurrentDistribution}, we conceptually split the current
into two components. One part is the circulating current surrounding
the moving vortex core as in the static case; we refer to this
component as the \emph{diamagnetic current}.  The other part which we
term the \emph{transport current} is the component which forces
vortices into motion.\footnote{Thinking of the system being embedded
  in three-dimensional space, the circular and transport currents are
  essentially the curl and gradient components of the current.}
      \begin{figure}[ht]
      \includegraphics[width=4.2cm]{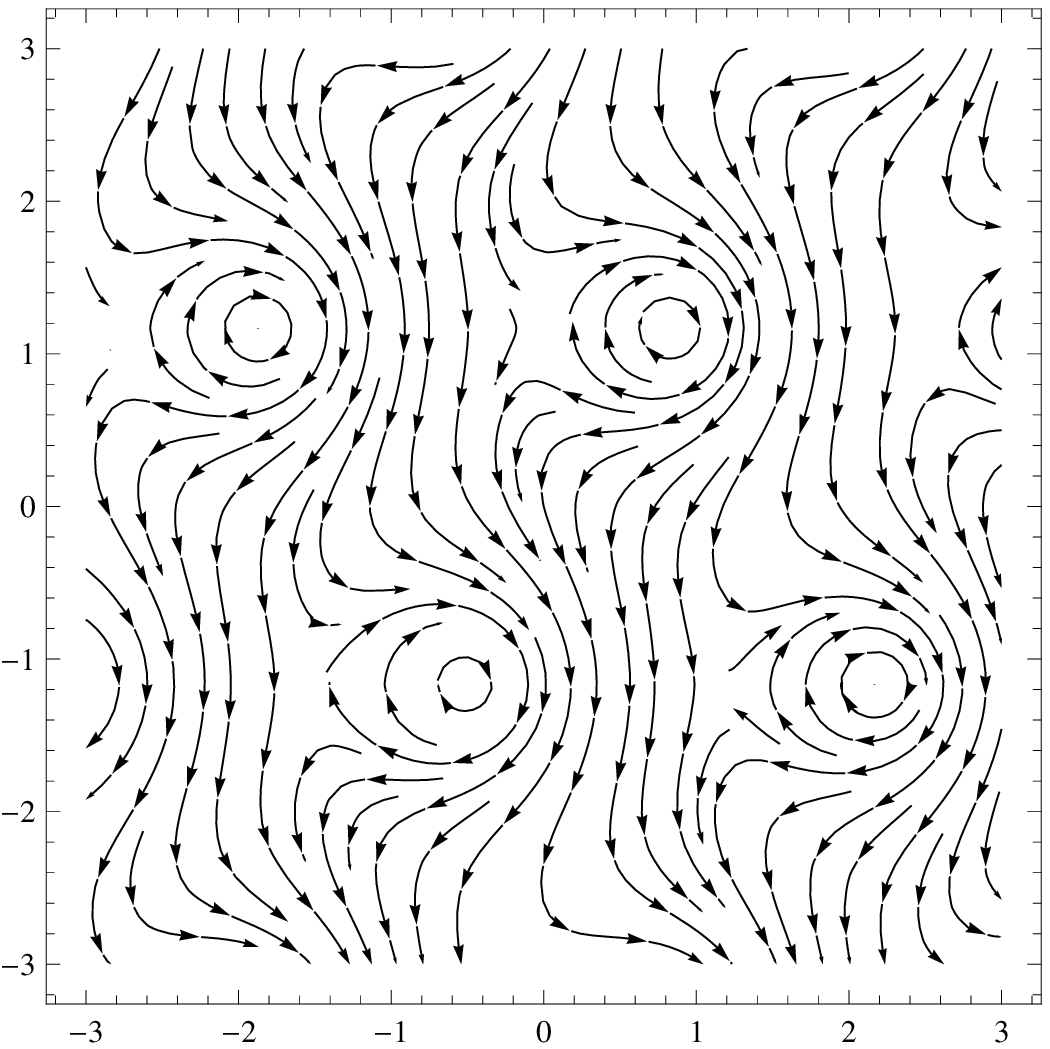}
      \includegraphics[width=4.2cm]{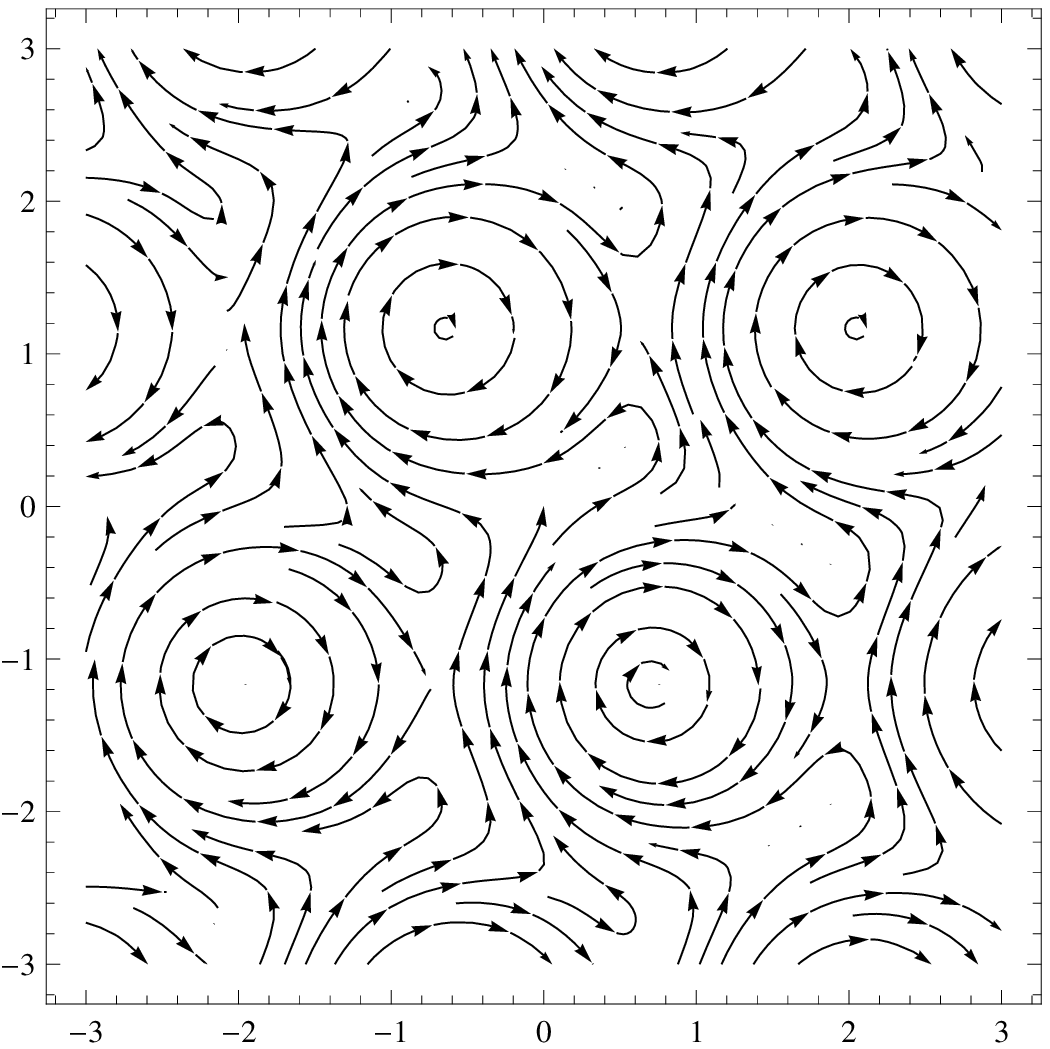}
    \caption{Current flow at $v=1,\omega=1$:
    (a) $t=0$; vortex cores move to the right.
    (b) $t=4\pi/5$; vortex cores move to the left. 
Vortices are drawn back and forth as the direction of the transport
current density alternates.  The magnitude of the current density has
maximal regions which tend to circumscribe the cores; the maxima in
these regions move in the plane and their manner of motion can be
described as leading the motion of the vortices by a small phase.  The
average current in a unit cell leads the motion of the vortex in time
by a phase of $\pi/2$.  }
    \label{CurrentDistribution}
    \end{figure}    

The diamagnetic current may be excised from our consideration by
integrating the current density over the unit cell $C$; that is,
we consider $\langle \mathcal J \rangle_{\bf r}$.  
We have $\langle {\mathcal J}_x\rangle_{\bf r}=0$ and 
    \begin{eqnarray}
    \label{avgj}
    \langle {\mathcal J}_y\rangle_{\bf r}
    &=&-\frac {\epsilon}{\beta_0}  {\tilde v} \cos(\omega \tau -\theta)
               e^{-\frac{{\tilde v}^2 }{4\omega}\sin2(\omega \tau-\theta)
               +\frac{v^2_{\omega}}{2}}
    .\end{eqnarray}
With our conventions, the transport current is along the
$y$-direction.
Considering the Lorentz force between the magnetic
flux in the vortices and the transport current, we expect the force on
the vortex lattice to be perpendicular.

We identify the locations of vortex cores to be where $|\Psi|^2=0$.
The velocity of the vortex cores turns out to be
    \begin{equation}
    \label{vv}
    v_c(\tau)={\tilde v}\cos ( \omega \tau-\theta)
    \end{equation}
along the $x$-direction. Note that vortex lattice moves coherently.
The vortex motion the electric field with a
phase $\theta$ which increases with frequency and reaches
$\pi/2$ asymptotically. The maximum velocity of vortex motion
$\tilde{v}$ decreases with increasing frequency.

In \citefg{CurrentDistribution} we show the current distribution
and the resultant oscillation of vortices.
As anticipated,  the transport current and the motion of
vortices \eqref{vv} are perpendicular as the vortices follow 
the input signal. The current density diminishes near the core; 
it is small there compared to its average value.

In steady-state motion, since the vortices move coherently in our approximation,
the interaction force between vortices is balanced as in the static case. 
Since the system is entirely dissipative, the motion that the vortices
collectively undergo is viscous flow. The vortex lattice responds to the 
Lorentz driving force as a damped oscillator, and this is the origin of
the frequency-dependent response.
\subsection{Configuration of moving vortex lattice}
\label{CML}
In static case the system is described by the GL equation.
Solving this equation, which is \eqref{1TDGL} but
with zero on the left-hand side, will select some lattice configuration. 
The global minima of the free energy correspond to a hexagonal lattice,
while there may be other configurations producing local minima.
In the static case the lattice configuration can be determined in 
practice by building an Ansatz from the linearized GL solution\cites{L65}
and then using a variational procedure to minimise the full free energy.

In the dynamic case, there is no free energy to minimise; we must embrace 
another method of making a physical prediction regarding the vortex lattice
configuration.  Let us follow \citep{LMR04} and take as the preferred
structure the one with highest heat-generation rate.  Though we have
at present no precise derivation, our physical justification of this
prescription is that the system driven out of equilibrium can reach
steady-state and stay in condensate only if the system can efficiently
dissipate the work done by the driving force. Therefore, whatever the
cause, the lattice structure most conducive to the maintenance of the
superconducting state will correspond to the maximal heat generation
rate.
 
The heat-generating rate\cites{KS98} is 
    \begin{eqnarray}
    \label{entropy}
   \langle  {\dot Q}  \rangle_{\bf r}  &=&2 \langle ~ |D_{\tau} \psi|^2\rangle_{\bf r} \\\nonumber
            &=&\frac{\epsilon}{\beta_0}\frac{{\tilde v}^2} {4}
    e^{\frac{{\tilde v}^2}{2} \cos^2 \theta-\frac{{\tilde v}^2}{4\omega}\sin2(\omega\tau-\phi)}\\\nonumber
    &&\big\{ \cos2(\omega\tau-\theta)+1+\frac{{\tilde v}^2}{8}\big[\cos4(\omega\tau-\theta)+1\big]
        \big\}
    .\end{eqnarray}
$\beta_0$ is given explictly in \eqref{beta0} and is the only parameter involving the apex
angle $\theta_A$ of the moving vortex lattice. Here $\beta_0$ plays 
the same r\^ole as the Abrikosov constant $\beta_A$ in the static case. 
 Corresponding to maximising the heat-generating rate, the preferred
 structure can be obtained by simply minimising $\beta_0$ with respect
 to 
$\theta_A$. This shows from the current viewpoint of maximal heat-generation 
rate that vortices are again expected to move coherently.

In \eqref{beta0} or \eqref{expbeta0} it is seen that the moving
lattice is distorted by the external electric field but this influence
subsides at high frequency. 
Numerical solution for minimising $\theta_A$ shows that while near the high-frequency limit there remain 
two local minima for $\beta_0$ with respect to $\theta_A$, the solution near 
$\theta_A=60\deg$ is favoured slightly over that at $30\deg$ as the global minimum.
This is as presented in \citefg{twovortexfigure}(a).
The two minima tend to approach each other slightly as the frequency begins to decrease further.
In an experimental setting, this provides an avenue for testing the 
empirical validity of the maximal heat generation prescription,
in particular in terms of the direction of lattice movement\cites{Fiory71}.

We put forth the physical interpretation
that at high frequency the friction force becomes less important, and
the distortion is lessened. 
Since interactions dominate the lattice structure the system at
high frequency will have many similarities with the static case.

\subsection{Energy dissipation in superconducting state}
\label{powerlose}

Energy supplied by the applied alternating current is absorbed and
dissipated by the vortex matter, and the heat generation does not
necessarily occur when and where the energy is first supplied.
In \citefg{TwoEnergyProfile}, we show an example of this transportation 
of energy by the condensate.
On the left is shown a contour plot of the work 
$\langle P \rangle_{\tau}=\langle{\mathcal
  J}\cdot {\bf v}\rangle_{\tau}$ done by the input signal; points
along a given contour are of equal power absorption.  On the
right of \citefg{TwoEnergyProfile} is shown the heat-generating rate\cites{KS98}, 
$\langle\dot Q\rangle_{\tau}=2\langle|D_{\tau}
\psi|^2\rangle_{\tau}$.  The periodic maximal regions are near the
vortex cores in both patterns.
     \begin{figure}[ht] 
      \includegraphics[width=4.2cm]{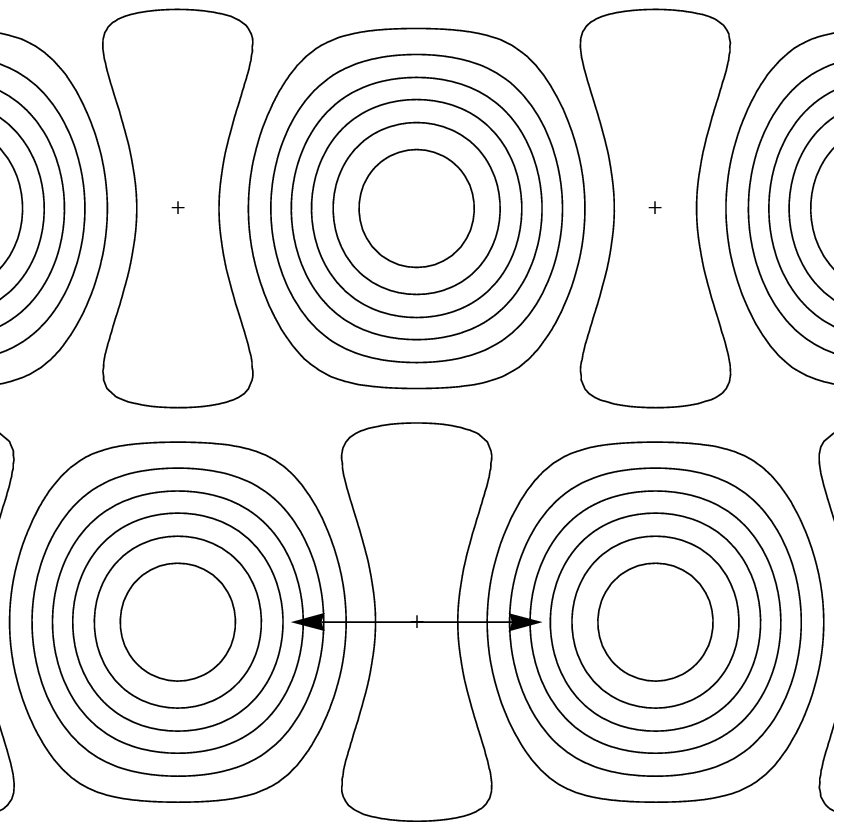}
      \includegraphics[width=4.2cm]{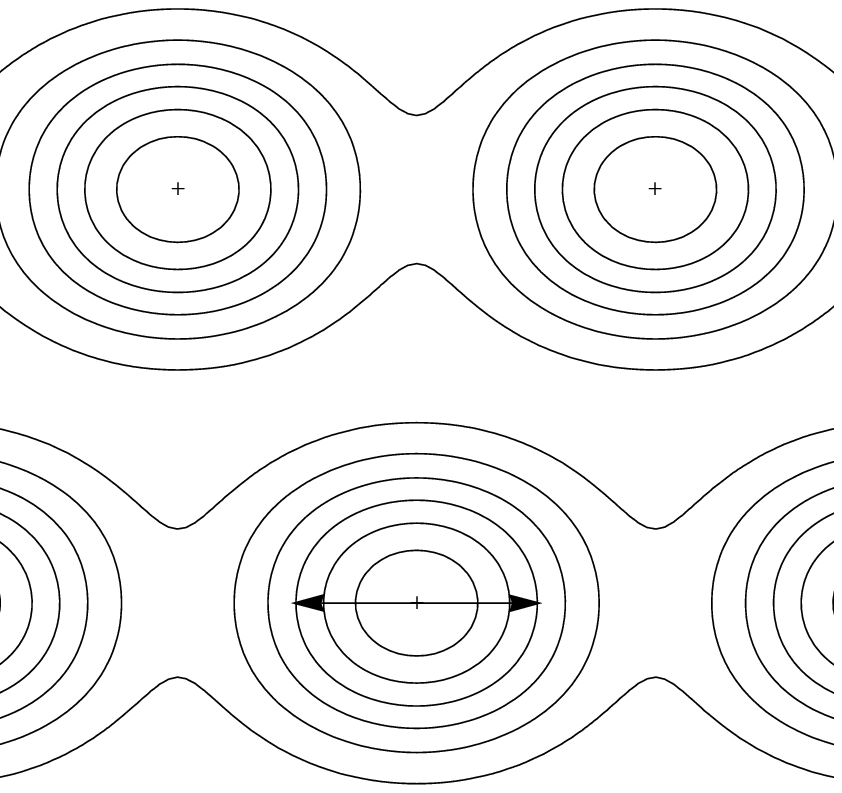}
    \caption{Work contours of superconducting component at $v=1$ and $\omega=1$:
    (a) Work $\langle P \rangle_{\tau}$ and
    (b) Heat generating rate $\langle \dot Q \rangle_{\tau}$.
    The vortex cores are denoted by `$+$' in both figures, shown in the $x$-$y$ plane.
    The maximum displacements of vortex cores are shown by the arrow. 
    The maximal region around the core in (a) is elongated by the current.
    The similar horizontal broadening around the core in (b) is caused by the vortex motion.
    Energy is transported; maxima in (a) and (b) do not coincide.
    }
    \label{TwoEnergyProfile}
    \end{figure}

In \citefg{TwoEnergyProfile}(b), one can see that the system dissipates energy via vortex cores.
From a microscopic point of view, Cooper pairs break into quasiparticles inside the core; 
these couple to the crystal lattice through phonons and impurities to transfer heat.
The interaction between vortices and excitation of vortex cores manifests
as friction\cites{Kpn02}. 

The power loss of the system averaged over time and space 
is $\langle P \rangle=\langle \dot Q \rangle$. 
    \begin{eqnarray}
    \langle P \rangle=\frac{\epsilon}{\beta_0}
    \frac{{\tilde v}^2}{2} e^{\frac{v_{\omega}^2}{2}}
            \bigg[
            I_0 \bigg( \frac{{\tilde v}^2}{4 \omega}\bigg)
            +\omega I_1\bigg( \frac{{\tilde v}^2}{4 \omega}\bigg)
            \bigg]
    \end{eqnarray} 
where $I_n$ is a Bessel function of the first kind.
      \begin{figure}[ht]
      \psfrag{avpow}{$\langle P \rangle$}
      \psfrag{eps}{$\epsilon$}
      \psfrag{nu}{$\nu$}
      \includegraphics[width=9cm]{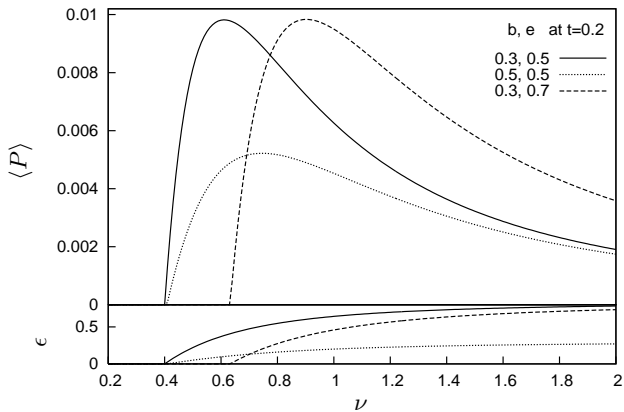}
    \caption{Power loss of supercurrent $\langle {\mathcal J}\cdot e\rangle$ (upper panel) 
    and expansion parameter $\epsilon$ (lower panel) as a function of frequency $\nu$.}
    \label{PowerLoss}
    \end{figure}
In \citefg{PowerLoss} is shown the power loss and also $\epsilon$ as a
function of frequency. $\epsilon$ is proportional the density of
Cooper pairs, and can be thought of as an indication of how robust is
the superconductivity.

As frequency increases, while $\epsilon$ tends to an asymptotic value,
$\langle P \rangle$ achieves a maximum and then decreases; this
maximum is due to fluctuations of order parameter caused by the input
signal.  In a fully dissipative system as considered here, the maximum
of each curve is not a resonance phenomenon but is instead caused by
fluctuation of the order parameter resulting from the influence of the
applied field.

A parallel may be drawn between what we have observed in this section
and the suppression of the superconductivity by macroscopic thermal
fluctuations commonly observed in high-temperature superconductors.
In our case, the vortices in a high-$T_c$ superconductor undergo
oscillation due to the driving force of the external field. We may think
of this as being analogous to the fluctuations of vortices due to
thermal effects alone in a low-$T_c$ superconductor. Although the
method of excitation is different, the external electromagnetic
perturbation in the present case essentially plays the same r\^ole as the
thermal fluctuations in low-$T_c$ situation.

Finally, we point out that $\epsilon$ seems to be an appropriate
parameter for determining the amount of power loss. Generically, it
seems that for points deeped inside the superconducting region, that
is at large $\epsilon$ compared with its saturation value at high 
$\nu$, the power loss due to the dissipative effects
of the vortex matter becomes suppressed.  We suggest the possibility
that this effect, which is na\"ively intuitive, is in fact physical
and more widely applicable than merely the present model.

\subsection{Generation of higher harmonics}

The practical application of superconducting materials
is dependent on how well one can control the inherent nonlinear behaviour.
In this section we will focus on the generation of higher harmonics in
the mixed state, in response to a single-frequency input signal.

The periodic transport current $\langle{\mathcal J}\rangle_{\bf r}$ is
an odd function of input signal, and it turns out that the response 
motion also contains only odd harmonics.
From \eqref{avgj} we can calculate
the Fourier expansion for transport current.
    \begin{equation} 
    \langle{\mathcal J}\rangle_{\bf r}=v {\rm Re} 
    		\big[ \sum_{n=0}^{\infty} 
    			\sigma^{(2n+1)} e^{i(2n+1)\omega\tau}
    		\big]
    ,\end{equation}
where the Fourier coefficient $\sigma^{(2n+1)}$ is
    \begin{equation}
    \label{hharmonics}
    \sigma^{(2n+1)}=
    \frac{\epsilon
         e^{\frac12v_{\omega}^2}
        i^n
        }
    {\beta_0\sqrt{1+\omega^2}}
     \bigg[
     iI_{n+1}\left( \frac{{\tilde v}^2}{4 \omega}\right)
     +I_n \left( \frac{{\tilde v}^2}{4 \omega}\right)
     \bigg]
    e^{-i(2n+1)\theta
    }
.\end{equation} 
We see the response goes beyond simple ohmic behaviour
and the coefficients are proportional to $\epsilon$.
Experimentally, one way of measuring these coefficients is a lock-in
technique\cites{lockin} which is adept at extracting a signal with a
known wave from even an extremely noisy environment.

To make contact with more standard parameters and satisfy our intuition, 
we expand the first two harmonics in terms of $v$. 
The fundamental harmonic, $\sigma^{(1)}$ expanded in powers of $v^2$ is
    \begin{eqnarray}
    \label{s1}
    \sigma^{(1)}&=&\frac{a_h}{\beta_A (1-i \omega)}\\\nonumber
    &&\mbox{}-
    \frac{a_h}{4 \beta_A (1-i \omega)}\frac{v^2}{1+\omega^2}\bigg(\frac{1-\beta_B/\beta_A}
    {1+\omega^2}+\frac1{a_h}+\frac{i}{2\omega}\bigg)\\\nonumber
    &&\mbox{}+{\mathcal O}(v^4)
   \end{eqnarray}
where $a_h=(1-t-b)/2b$. 
The first term is the ohmic conductivity denoted as $\sigma_0^{(1)}$, 
and is reminiscent of Drude conductivity for free charged particles. 
This is not an unexpected parallel, since the Cooper pairs in a 
superconducting system can be imagined to behave like a free-particle gas.
Taking this viewpoint, in the small-signal limit, the ratio
$\imag \sigma / \real \sigma=\omega=\Omega \tau_s$ gives the relaxation time of the 
charged particles.
Subsequent higher-order corrections all contain $\omega$ 
in such a way that their contributions are suppressed at large $\omega$.
The coefficient of the $n=1$ harmonic expanded in powers of $v^2$ is 
    \begin{equation}
    \label{s3}
    \sigma^{(3)}=\frac {a_h}{8\beta_A} \frac {v^2/\omega}{\omega(3-\omega^2)-i(3\omega^2-1)}
    +{\mathcal O}(v^4)
    \end{equation}
which decreases quickly with increasing $\omega$.

 In \citefg{GeneratingHarmonics}, we show the generation of higher harmonics 
 for three different states in the dynamical phase diagram.
For each harmonic labeled by $n$, $|\sigma^{(2n+1)}|$ as a function 
of $\nu$ has the same onset as $\epsilon$.
We can see that $|\sigma^{(2n+1)}|$ reaches a maximum and then starts 
to decay while $\epsilon$ saturates.
The coefficients of harmonics with $n>0$ decay to zero in the high $\nu$ limit,
where the state is well inside the superconducting region.

We pointed out in \secref{powerlose} and reaffirm here that $\epsilon$ plays a significant 
r\^ole in determining the extent of nonlinearity in the system.
In turn, the parameter which controls this is $\omega$. When $\omega$ is large,
$\epsilon$ is brought closer to its saturation value $\epsilon_\infty$, 
causing the higher harmonics to be suppressed,
and also lessening distortion of the vortex lattice.
Finally, for a given harmonic, $|\sigma^{(2n+1)}|$ is generally smaller when 
$\epsilon_\infty$ is smaller; this can be seen by comparing (a) and 
(c) of \citefg{GeneratingHarmonics}.
One might point out that the nonlinear behaviour is decreased at, for example, large $\omega$.
Nevertheless, we view the parameter $\epsilon/\epsilon_\infty$ as more intrinsic to the system,
rather than simply characterising the input signal.

A limited parallel can be drawn between the effect of thermal noise in high-$T_c$
superconducting systems, and the effect of the electromagnetic perturbation in our present case.
It seems that in either case the fluctuation influence can be reduced by moving the state 
deeper inside the superconducting region.
    \begin{figure*}[ht!]
	\psfrag{si1}{$\sigma^{(1)}$}
	\psfrag{si3}{$\sigma^{(3)}$}
	\psfrag{si5}{$\sigma^{(5)}$}
	\psfrag{Abs}[r][r]{$|\sigma|$}
	\psfrag{ResS}[r][r]{$\real\sigma$}
	\psfrag{ImsS}[r][r]{$\imag\sigma$}    
	\psfrag{mImsS}[r][r]{$-\imag\sigma$}
	\psfrag{nu}{$\nu$}
	\psfrag{eps}{$\epsilon$}
	\psfrag{aaa}{(a)}
	\psfrag{bbb}{(b)}
	\psfrag{ccc}{(c)}
    \includegraphics[width=18cm]{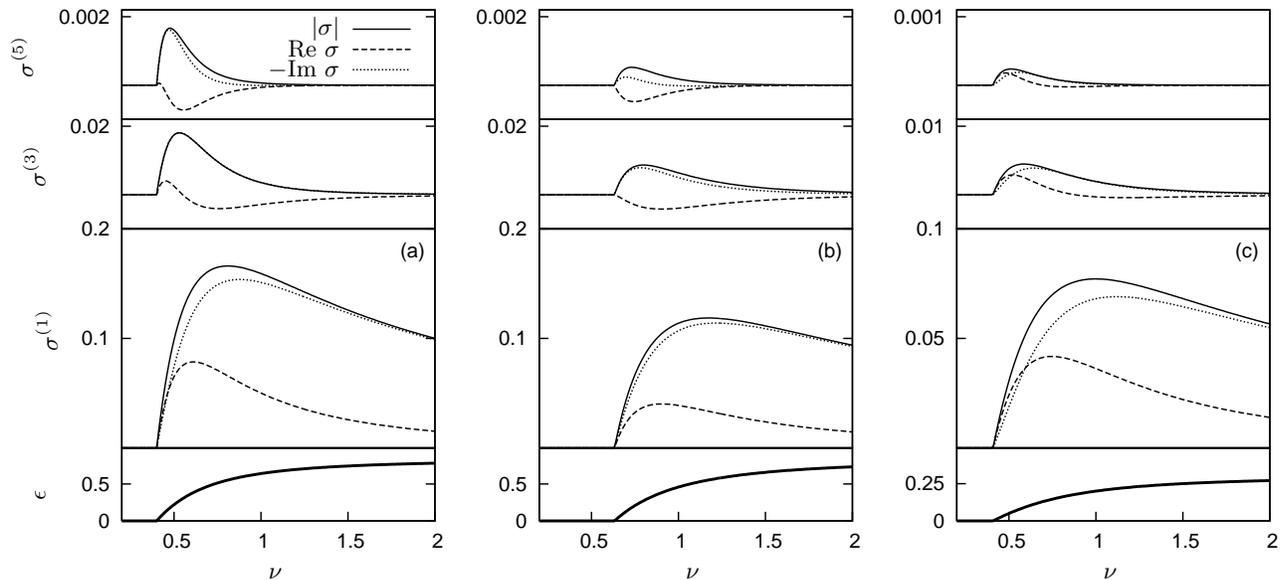}
    \caption{Generation of harmonics: $\sigma^{(1)}$, $\sigma^{(3)}$ 
      and $\sigma^{(5)}$ are plotted with respect to frequency $\nu$ at $t=0.2$.
    (a) $b=0.3$, $e=0.5$. (b) $b=0.3$, $e=0.7$. (c) $b=0.5$, $e=0.5$.
    The distance from the dynamic phase transition boundary $\epsilon$ is shown in the 
    separate bottom panel. The scale of figure (c) is half that of (a) and (b). 
    }
    \label{GeneratingHarmonics}
    \end{figure*}
\subsection{Flux-flow Hall effect}
\label{halleffect}
In contrast to the fully dissipative system we have considered, in
this section we will discuss an effect caused by the non-dissipative
component, namely the Hall effect.  In a clean system, vortices move
without dissipation; a transverse electric field with respect to
current appears.  The non-dissipative part is subject to a
Gross-Pitaevskii description, using a type of nonlinear Schr\"odinger   
equation\cites{Kpn02}, with a non-dissipative part to the relaxation
$\gamma$ from \eqref{1TDGL}.  The
fully dissipative operator $L$ in our previous discussion 
can be generalised by using a complex relaxation
coefficient $r=1+i\zeta$. We thus define
    \begin{equation}
    {\mathcal L}=r D_{\tau}-\frac12(D^2_x+\partial_y^2+\partial_z^2)
    .\end{equation}
The ratio $\zeta=\imag \gamma /\real \gamma$ is typically on the order
of $10^{-3}$ for a conventional superconductor, and $10^{-2}$ for a
high-$T_c$ superconductor\cites{Kpn02}.

The Hall Effect is small here. In normal metals, the non-dissipative
part gives the cyclotron frequency.  If $\tau_e$ is the relaxation
time of a free electron in a dirty metal, then for typical values of
$\omega_c \tau_e\ll1$ the Hall effect becomes negligible.  Because the
supply of conducting electrons is limited, the transverse component
increases at the expense of the longitudinal component as the mean
free path of excitations grows.  It is equivalent to an increase in
the imaginary part of the relaxation constant at the expense of the
real part.

The Eigenvalues and Eigenfunctions of ${\mathcal L}$ can be obtained
easily by replacing the $v$ in previous results with $r v$, $\omega$
with $r \omega$ and $\tau$ with $\tau/r$.  The transport current along
the $x$-direction is no longer zero in the presence of the
non-dissipative component; it is propotional to $\zeta$.  The
frequency-dependent Hall conductivity can be obtained from the
first-order expension in $v$,
    \begin{eqnarray}
    \label{sh}
    \sigma^{h(1)}_{0}=\frac{a_h}{\beta_A}\frac{\zeta}{(1-i\omega)^2-\omega^2\zeta^2}
    \end{eqnarray}
while the Hall contribution in the $y$ direction is expected to be
negligible, as it is of the order of $\zeta^2$.
   
In principle, the crossover between non-dissipative systems and
dissipative systems can be tuned using the ratio $\zeta$.  In a
non-dissipative system, which is the clean limit, the Hall effect is
important and taking account of the imaginary part of TDGL is
necessary.  On the contrary, in a strong dissipative system where
excitations are in thermal equilibrium via scattering, the TDGL
equation gives satisfactory agreement.

\section{Experimental Comparison}
\label{Discussion} 
Far-Infrared spectroscopy can be performed using monochromatic
radiation which is pulsed at a high rate, known as Fast Far-Infrared
Spectroscopy.  This technique sports the advantage of avoiding
overheating in the system, making it a very effective tool in observing
the dynamical response of vortices.  In particular, one can study the
imaginary part of conductivity contributed mainly from superconducting
component.

In \citefg{FIR} is shown a comparison with an NbN experiment measuring
the imaginary part of conductivity.
The sample has the gap energy $2\Delta=5.3$~meV.
The resulting value of $2\Delta/T_c$ is larger than the value expected from
BCS theory\cites{Ikb09}.
We consider frequency-dependent conductivity in the case of linearly
polarised incident light with a uniform magnetic field along the $z$ axis.  

The theoretical conductivity contains both a superconducting and a normal contribution.
The total conductivity is obtained from the total current as in
\eqref{Amplaw} where the normal-part conductivity in the condenstate
is the conductivity appearing in the Drude model.

According to our previous discussion, the nonlinear effect of the
input signal on NbN is unimportant in the THz region, which corresponds to $\nu\sim17$.
An approximation where the flux-flow conductivity includes only the term
$\sigma^{(1)}_0$ from \eqref{s1}, and Hall coefficient
$\sigma^{h(1)}_0$ from \eqref{sh} is shown in \citefg{FIR} and the
agreement with experiment is good.
   	
The na\"ive way in which we have treated the normal-part contribution
is essentially inapplicable to the real-part conductivity.  This is
because the real-part conductivity contains information about
interactions with the quasi-particles inside the core, making further
consideration necessary\cites{LL09}.
\begin{figure}[ht] 
\psfrag{imagsigunits}{$\imag \sigma$ [$10^3$/$\Omega$cm]}
\psfrag{omegamev}{$\omega$ [meV]}
\psfrag{sigadded}[r][r]{$\sigma_n+\sigma$} 
\psfrag{3T}{3~T}
\psfrag{5T}{5~T}
\psfrag{7T}{7~T}
\includegraphics[width=6cm,angle=270]{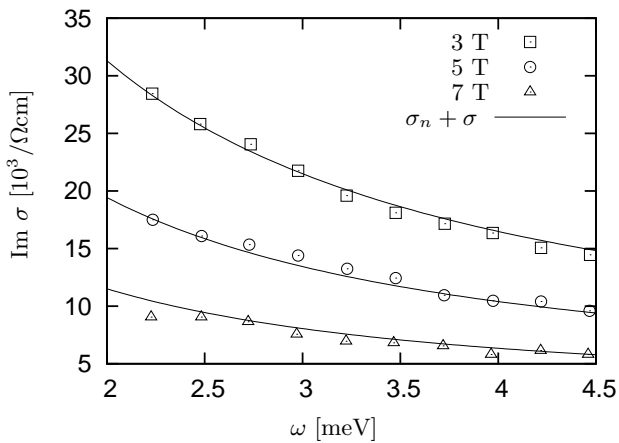}
\caption{Experimental comparison of imaginary part of conductivity at
  high frequency:
  The NbN experimental data are from \litfig{6(b)} of \citep{Ikb09}. 
  Material parameters $\Tnst=15.3$~K and $H^0_{c2}=14.1$~T calculated
  using \litfig{3(b)} of \citep{Ikb09}.
  Normal-state conductivity is $\sigma_n=20\cdot10^4 (\mu \Omega$~cm$)^{-1}$ 
  and relaxation time of an electron $\tau_e=5$~fs are taken from \citep{Ikb09}. 
  The theoretical curve has one fitting parameter $\kappa=44.5$.}
\label{FIR}
\end{figure}

\section{Conclusion}
\label{Conclusion}
The time-dependent Ginzburg-Landau equation
has been solved analytically to study the dynamical response of the free vortex lattice.  
Based on the bifurcation method, which involves an expansion in the distance to the phase
transition boundary, we obtained a perturbative solution to all
orders.  We studied the response of the vortex lattice in the
flux-flow region just below the phase transition, at first order in
this expansion.
We have seen that there are certain parameters which can  
be tuned using the applied field and temperature, providing a feasible 
superconducting system where one can study precise control of nonlinear 
phenomena in vortex matter.

Under a perturbation by electromagnetic waves, the steady-state
solution shows that there is a diamagnetic current circulating the
vortex core, and a transport current parallel to the external electric
field with a frequency-dependent phase shift and amplitude.  Vortices
move perpendicularly to the transport current and coherently.

Using a technique of maximising the heat-generation rate, we showed
that the preferred structure based on energy dissipation is
a hexagonal lattice, with a certain level of distortion appearing
as the signal is increased or the frequency is lowered.

Energy flowing into the system via the applied field is dissipated
through the vortex cores.  
We showed that the superconducting part may be thought of as having 
inductance in space and time.

We have written transport current 
beyond a simple linear
expression.  A comparison between different harmonics of three 
different states in our four-dimensional parameter space 
indicated that the nonlinearity becomes unimportant at
high frequency and small amplitude, and the influence of the input
signal is decreased when the system moves deeper inside the
superconducting region, away from the phase-transition boundary.

To observe the configuration of moving vortices, techniques such as
muon-spin rotation\cites{Fg02,Fg06}, SANs\cites{Fg06}, STM\cites{stm} and others\cites{AbrikosovTechniques} 
seem to be promising options. 
To provide the kind of input signal considered here, methods such as 
short-pulse FIR Spectroscopy as used in \citep{Ikb09} might be applied. 
The coefficient we defined
in \eqref{hharmonics} corresponds to conductivity.  We have also seen
that a simple parametrisation by complex quantities like conductivity
and surface impedence is insufficient to capture the detailed
behaviour of the system; in performing experiments, it should be kept
in mind that the nonlinearity can be measured in terms of more
appropriate variables as we have shown. 

We have viewed the forcing of the system by the applied field to be
somewhat analogous to thermal fluctuations, in the sense that they
both result in vibration of the vortex lattice.  Hence, the influence
of the electromagnetic fluctuation is stronger at the nucleation
region of superconductivity than deep inside the superconducting
phase.  Besides, since at high frequency the motion of vortices is
limited, the influence from electric field is suppressed, as is the
Hall effect.
\bigskip
\begin{acknowledgments}
Fruitful discussions with B.~Rosenstein, V.~Zhuravlev, and J.~Kol{\'a}{\v c}ek 
are greatly appreciated. 
The authors kindly thank G.~Bel and P.~Lipavsk{\'y} for critical reading of 
the manuscript and many useful comments.
The authors also have benefitted from comments of A.~Gurevich.
We thank J.~R.~Clem for pointing out to us reference \citep{Fiory71}.

\texttt{NSC99-2911-I-216-001}
\end{acknowledgments}

\appendix 
\section{TDGL parameters}
\label{Gamma}
We follow \citep{Kpn02} to estimate the coefficient $\gamma$ which
characterises the relaxation process of the order parameter. $\gamma$
is the inverse of the diffusion coefficient for electrons in the normal state.
For a strongly-scattering system, as in the dirty limit\cites{Kpn02}, 
the ratio between the relaxation times of order
parameter
\begin{equation}
\tau_{\Delta}=\frac{\hbar^2\gamma}{2 m_{ab}^*|\alpha \Tnst(1-t)|}
\end{equation}
and the vector potential (or current)
\begin{equation}
\tau_j=\frac{\beta \sigma_n 2 m_{ab}^*}{8 e^2|\alpha \Tnst(1-t)|}
\end{equation}
is
\begin{equation}
\frac{\tau_{\Delta}}{\tau_j}=\frac{\pi^4}{14\zeta(3)}
.\end{equation}
By definition of the thermal critical field $H_c$ and $H_{c2}=\sqrt{2}
\kappa H_c$, as in \citep{Tinkham}, we know the ratio of the two
parameters is
\begin{equation}
\frac{(\alpha \Tnst)^2}{\beta}=\frac{(H^0_{c2})^2}{8\pi^2\kappa^2}.
\end{equation}
The coherence length at zero temperature can be written
in terms of $H^0_{c2}$ as $\xi^2=\Phi_0/2\pi H^0_{c2}$ and
in terms of effective mass as $\xi^2=\hbar^2/2m^*\alpha \Tnst$.
As a result,
\begin{equation}
\gamma=\frac{2 \pi^5 \kappa^2 \sigma_n}{7 \zeta(3) c^2}
.\end{equation}
With $\gamma$, we can retrieve the experimental quantities from
the rescaled ones used in calculation.
Using the $0$-subscripted original variables, we write
the electric field $E=\big(2 \hbar/e^*\xi^3\gamma\big)e$, and
the frequency $\Omega=\big( 2/\gamma \xi^2\big)\nu$.
The current density ${\bf J}_0 = (c H^3/\sqrt{2\pi \Phi_0} \kappa^2)^{-1} {\bf J}$.
In the case of linear response, we have 
${\bf J}_0=\sigma_0 {\bf E}$ where 
$\sigma_0=\big(c^2 \gamma/4\pi\kappa^2\big)\sigma$.

\section{Solving the linearised TDGL equation}
\label{LTDGL_sol}
We consider the linearised time-dependent Ginzburg-Landau equation 
\eqref{LTDGL}, which has been written in our chosen gauge.  
We wish to find the set of Eigenfunctions of $L$ corresponding to the lowest
Eigenvalue. Based on knowledge of the solution in the static case,
we solve the now time-dependent problem by making the following Ansatz\cites{Lo93,BRpc}.
The electric field along the $y$-direction breaks rotational symmetry
in the $x$-$y$ plane, so we write 
\begin{equation}
\label{an}
f(x,y,z,\tau) = e^{ik_z z}e^{ik_x x}e^{g_2(\tau)y^2+g_1(\tau) y+g_0(\tau)}
.\end{equation}

After substitution of $f$ for $\Phi$ in \eqref{LTDGL}, comparison of coefficients of powers of $y$ gives
the following differential equations in $\tau$.
\begin{eqnarray}
\label{geq}
\dot {g}_2+\frac12-2 g_2^2&=&0, \\
\dot {g}_1+i v \cos \omega \tau-k_x-2g_2 g_1&=&0, \\
\dot {g}_0+\frac12(k_x^2+k_z^2)-\frac12 g_1^2-g_2&=&\varepsilon
.\end{eqnarray}
The solutions are
\begin{equation}
g_2=-\frac12\tanh[c_2+\tau]
.\end{equation}
For a steady-state solution, $\tau\rightarrow\infty$, we have $g_2=-\frac12$. 
\begin{eqnarray}
g_1=c_2 e^{-\tau}+k_x-i \tilde v \cos(\omega \tau-\theta)
.\end{eqnarray} 
As with $g_1$, we have here $c_2=0$. 
\begin{equation}
g_0= - ik_x\frac{\tilde v}{\omega}\sin(\omega\tau-\theta)
 -\frac{{\tilde v}^2}{8\omega}\sin{2(\omega\tau-\theta)}+c_0 
\end{equation}
where $c_0$ is a normalisation constant. 

The resulting Eigenvalue is
\begin{eqnarray}
\label{Leigenv}
\varepsilon=\frac {k_z^2}2+\frac12+\frac{v^2}{4(1+\omega^2)}.
\end{eqnarray}    
Now, although in a more realistic treatment of the system, one may introduce some boundary condition
restricting $k_z$ to a certain set of values, here we simply select the smallest Eigenvalue $\varepsilon_0$
available to us by setting $k_z$ to zero.
Thus equipped with the set of Eigenfunctions corresponding to our lowest Eigenvalue,
we deem them to be the first elements of our basis, labeled by $n=0$ and $k_x$.
These Eigenfunctions of $L$ are 
 \begin{equation}
    \label{appdsol}
      \varphi_{0,k_x}(x,y,z,\tau)=
                    e^{i k_x (x- v \sin\omega \tau/\omega)} \tilde {u}_{k_x}(y,\tau)
    \end{equation}
    with
    \begin{eqnarray}
    	u_{k}(y,\tau)=c(\tau)e^{  -\frac12
    	         \big[
                    y-k_x+i {\tilde v} \cos(\omega \tau-\theta)
                \big]^2
                }
    ,\end{eqnarray}
    and
    \begin{eqnarray}
    \label{appct}
    c(\tau)=e^{
            -\frac{{\tilde v}^2}{4}
                \big[
                    \sin^2\theta+
                    \cos 2(\omega \tau-\theta)+
                    \frac1{2\omega}\sin2(\omega \tau-\theta)
                \big]
            }
.\end{eqnarray}

In the same way, the corresponding Eigenfunctions of $L^{\dagger}$ can be obtained.
    \begin{equation}
    \label{dsol}
        \tilde{\varphi}_{0,k_x}(x,y,z,\tau)=e^{i k_x (x- v \sin\omega \tau/\omega)} \tilde {u}_{k_x}(y,\tau)
    \end{equation}  with
    \begin{eqnarray}
    \label{dukv}
    \tilde{u}_{k_x}(y,\tau)=\tilde {c}(\tau)e^{  -\frac12
                \big(
                    y-k_x-i {\tilde v} \cos(\omega \tau+\theta)
                \big)^2
                },
    \end{eqnarray}
    where
    \begin{eqnarray}
    \label{cd}
    \tilde {c}(\tau)=e^{
            -\frac{{\tilde v}^2}{4}
                \big[
                    \sin^2\theta+
                    \cos 2(\omega \tau+\theta)-
                    \frac1{2\omega}\sin2(\omega \tau+\theta)) 
                \big]
            }
    \end{eqnarray}
according to our normalisation condition \eqref{normalised}.  The
lowest Eigenvalue of $L^{\dagger}$ is $\tilde{\varepsilon}_0=\varepsilon_0$.

In the $\omega \rightarrow 0$ limit, the system reduces to the case of
constant electric field. The Eigenfunctions and Eigenvalues are then
consistent with those obtained by Hu and Thomson\cites{HT71}.
In the limit of zero electric field, the Eigenfunctions and
Eigenvalues reduce to those of the lowest Landau level static-state
solution\cites{Tinkham}.  This is also the same in the
$\omega\rightarrow \infty$ limit.


\begin{thebibliography}{99} 
\bibitem{O07} D.~E.~Oates, \emph{Overview of Nonlinearity in HTS: What We Have Learned and Prospects for Improvement}, J.~Supercond.~Novel Magn.~\textbf{20}, 3 (2007); J.~C.~Booth, S.~A.~Schima, D.~C.~DeGroot, \emph{Description of the Nonlinear Behavior of Superconductors Using a Complex Conductivty}, IEEE Trans.~Appl.~Supercond.~\textbf{13}, 315 (2003); A.~V.~Velichoko, M.~Courier, J.~Lancaster, A.~Porch, \emph{Nonlinear microwave properties of high $T_c$ thin films}, Supercond.~Sci.~Technol.~\textbf{18} R24 (2005)
\bibitem{Tinkham} M.~Tinkham, \emph{Introduction to Superconductivity}, McGraw-Hill, New York (1996)
\bibitem{Fg06} D.~Charalambous, E.~M.~Forgan, S.~Ramos, S.~P.~Brown, R.~J.~Lycett, D.~H.~Ucko, A.~J.~Drew, S.~L.~Lee, D.~Fort, A.~Amato, U.~Zimmerman, \emph{Driven vortices in type-II superconductors: A muon spin rotation study}, Phys.~Rev.~B \textbf{73}, 104514 (2006)
\bibitem{RL09} B.~Rosenstein, D.~Li, \emph{Ginzburg-Landau theory of type II superconductors in magnetic field}, Rev.~Mod.~Phys.~\textbf{82}, 109 (2010) and references therein
\bibitem{Ab57} A.~A.~Abrikosov, Zh.~Eksperim, i Teor.~Fiz.~\textbf{32}, 1442 (1957)
\bibitem{GB97} A.~Gurevich, E.~H.~Brandt, \emph{AC response of thin superconductors in the flux-creep regime}, Phys.~Rev.~B \textbf{55}, 12706 (1997)
\bibitem{Kpn02} 
N.~B.~Kopnin, \emph{Theory of nonequilibrium superconductivity}, University Press, Oxford (2001)
\bibitem{CC91} M.~W.~Coffey, J.~R.~Clem, \emph{Unified theory of effects of vortex pinning and flux creep upon the RF surface impedance of type-II superconductors}, Phys.~Rev.~Lett.~\textbf{67}, 386 (1991)
\bibitem{GR66} J.~I.~Gittleman, B.~Rosenblum, \emph{Radio-Frequency Resistance in the Mixed State for Subcritical Currents}, Phys.~Rev.~Lett.~\textbf{16}, 734 (1966)
\bibitem{HT71} C.-R.~Hu, R.~S.~Thompson, \emph{Dynamic Structure of Vortices in Superconductors. II. $H\ll H_{c2}$}, Phys.~Rev.~B \textbf{6}, 110 (1972); R.~S.~Thompson, C.-R.~Hu, \emph{Dynamic Structure of Vortices in Superconductors}, Phys.~Rev.~Lett.~\textbf{27}, 1352 (1971)
\bibitem{TDGLnumerical} G.~W.~Crabtree, D.~O.~Gunter, H.~G.~Kaper, A.~E~Koshelve, G.~K. Leaf, V.~M.~Vinokur, \emph{Numerical simulation of driven vortex systems}, Phys.~Rev.~B \textbf{61}, 1146 (2000); D.~Y.~Vodolazov, F.~M.~Peeters, \emph{ Rearrangement of the vortex lattice due to instabilities of vortex flow}, Phys.~Rev.~B \textbf{76}, 014521 (2007)
\bibitem{LMR04} D.~Li, A.~Malkin, B.~Rosenstein, \emph{Structure and orientation of the moving vortex lattice in clean type-II superconductors}, Phys.~Rev.~B \textbf{70}, 214529 (2004)
\bibitem{ZR07} 
B.~Rosenstein and V.~Zhuravlev, \emph{Quantitative theory of transport in vortex matter of type-II superconductors in the presence of random pinning}, Phys.~Rev.~B \textbf{76}, 014507 (2007)
\bibitem{KS98} J.~B.~Ketterson, S.~N.~Song, \emph{Superconductivity}, University Press, Cambridge (1998)
\bibitem{GE68} L.~L.Gor'kov, G.~M.~Eliashberg, Zh.~Eksp.~Teor.~Fiz. \textbf{54}, 612 (1968); A.~Schmid, Phys.~Kond.~Materie \textbf{5}, 302 (1966)
\bibitem{L65} G.~Lasher, \emph{Series Solution of the Ginzburg-Landau Equations for the Abrikosov Mixed State}, Phys.~Rev.~\textbf{140}, A523 (1965)
\bibitem{LR99} D.~Li, B.~Rosenstein, \emph{Lowest Landau level approximation in strongly type-II superconductors}, Phys.~Rev.~B \textbf{60}, 9704 (1999)
\bibitem{Kim99} P.~Kim, Z.~Yao, C.~A.~Bolle, C.~M.~Lieber, \emph{Structure of flux line lattices with weak disorder at large length scales}, Phys.~Rev.~B \textbf{60}, R12589 (1999)
\bibitem{Fg02} D.~Charalambous, P.~G.~Kealey, E.~M.~Forgan, T.~M.~Riseman, M.~W.~Long, C.~Goupil, R.~Khasanov, D.~Fort, P.~J.~King, S.~L.~Lee, F.~Ogrin, \emph{Vortex motion in type-II superconductors probed by muon spin rotation and small-angle neutron scattering}, Phys.~Rev.~B \textbf{66}, 054506 (2002)
\bibitem{Fiory71} A.~T.~Fiory, \emph{Quantum interference effects of a moving vortex lattice in Al films}, Phys.~Rev.~Lett.~\textbf{27}, 501 (1971)
\bibitem{lockin} M.~O.~Sonnaillon, F.~J.~Bonetto, \emph{A low-cost, high-performance, digital signal processor-based lock-in amplifier capable of measuring multiple frequency sweeps simultaneously}, Rev.~Sci.~Inst.~\textbf{76}, 024703 (2005)
\bibitem{Ikb09} Y.~Ikebe, R.~Shimano, M.~Ikeda, T.~Fukumura, M.~Kawasaki, \emph{Vortex dynamics in a NbN film studied by terahertz spectroscopy}, Phys.~Rev.~B \textbf{79}, 174525 (2009)
\bibitem{LL09} P.-J.~Lin, P.~Lipavsk\'{y}, \emph{Time-dependent Ginzburg-Landau theory with floating nucleation kernel: Far-infrared conductivity in the Abrikosov vortex lattice state of a type-II superconductor}, Phys.~Rev.~B \textbf{80}, 212506 (2009)
\bibitem{stm} A.~M.~Troyanovski, J.~Aarts, P.~H.~Kes, \emph{Collective and plastic vortex motion in superconductors at high flux density}, Nature, \textbf{339},665(1999)
\bibitem{AbrikosovTechniques} T.~H.~Johansen, \emph{Gallery of Abrikosov Lattices in Superconductors}, \texttt{http://www.fys.uio.no/super/vortex}
\bibitem{Lo93} C.~F.~Lo, \emph{Propagator of the general driven time-dependent oscillator}, Phys.~Rev.~A \textbf{47}, 115 (1993)
\bibitem{BRpc} B.~Rosenstein, private communication, unpublished.
\end{thebibliography}
\end{document}